\documentclass[letter]{aa}  

\usepackage{graphicx}
\usepackage{txfonts}
\begin{document}

   \title{Preliminary evidence for a Universal Magnetic Field Configuration in Disk-Galaxy Halos with POSSUM}
   \titlerunning{A Universal Magnetic Field Configuration}


   \author{Ioannis Contopoulos\inst{1}
	\thanks{\email{icontop@academyofathens.gr}}
		\and
		Ioannis Myserlis\inst{2}
          \and
          Shane O'Sullivan\inst{3}
		\and
        Sunil Malik\inst{3}
        \and
		Takuya Akahori\inst{4}
          }

    \authorrunning{Contopoulos et al.}

   \institute{ Research Center for Astronomy and Applied Mathematics, Academy of Athens, Athens 11527, Greece
				\and
Instituto de Radioastronom\'{i}a Milim\'{e}trica, E-18012 Granada, Spain
				\and
Departamento de Física de la Tierra y Astrofísica \& IPARCOS-UCM, Universidad Complutense de Madrid, 28040 Madrid, Spain
				\and
Mizusawa VLBI Observatory, National Astronomical Observatory of Japan, 2-21-1 Mitaka, Tokyo 181-8588, Japan
             }

   \date{Received; Accepted}


\abstract{
Recent LOFAR studies that exploited rotation measures (RMs) of background polarized sources have begun to trace the magnetic field environment of galactic halos, revealing a gradual decline of field strength with height above the disk that is consistent with expectations from magnetized galactic winds and outflows.
}{
In this paper we use the POSSUM catalog with its much higher density of background RM sources (30-50 per square degree) to extend this work and further  investigate the sign of the halo magnetic field. We are searching for the universal configuration that arises if halo magnetization originates from a central Cosmic Battery, and is subsequently transported outward by large-scale outflows.
}{
Our analysis considers 10 edge-on galaxies in the field of view of the POSSUM survey with angular sizes between $6'$ and $10'$. For each such galaxy we collect all available background RM sources in a circular region with a radius of $2^\circ$ around it. We rescale all galaxy sizes and their associated background RM source distances to a common galaxy size of $10'$ and stack all these regions together. 
We divide the stacked configuration of RM sources
into the four quadrants on either side of the stacked galaxy axis. The mean RM of background sources is computed in each quadrant, and we obtain a pattern of positive and negative mean RM values around the stacked galaxy.
}{
A consistent, systematic RM-gradient pattern emerges inside about 6 stacked galactic radii, as expected in Magnetic Disk-Winds with the axial magnetic field configuration predicted by the Cosmic Battery. The inferred large-scale axial electric current in the halo flows from the center outwards. Beyond 6 galactic radii we observe the reverse mean RM pattern, as expected in the axial Magnetic Black-Hole Jet configuration predicted by the Cosmic Battery..
}{
Our work confirms the prediction of the Cosmic Battery, and presents one of the very few known signatures of the magnetic field in the halo and the circumgalactic medium (CGM).
}


   \maketitle

\label{firstpage}

\section{Scope}

A new generation of broadband radio polarimetric facilities such as LOFAR, ngVLA, ALMA, and the SKA now allows magnetic fields along the line of sight to be probed with unprecedented detail.  Large-scale ordered fields are central to key astrophysical processes, from star
formation and cosmic-ray confinement to accretion dynamics and jet launching. Understanding their origin and long-term evolution remains one of the major unsolved problems in astrophysics.

Linearly polarized emission offers a powerful diagnostic of magnetized astrophysical plasma mediums.
%
Propagation through a magnetized plasma rotates the electric-vector position angle $\chi$ (EVPA) according to $\chi = \chi_0 + \mathrm{RM} \,\lambda^2$, with the Faraday Rotation Measure (RM) defined as
\begin{equation}
   \mathrm{RM} =
   \frac{e^3}{2\pi m_e^2 c^4}\,
   \int_{r_0}^{0} n_e\, B_{\rm l.o.s.}\, dr.
\end{equation}
Here, $\lambda$ is the electromagnetic wave's wavelength, $n_e$ and $B_{\rm l.o.s.}$ are the electron density and the magnetic field component along the line of sight, respectively, and $r_0$ is the distance to the emitting source.  Thus both the magnitude and sign of RM encode information about the strength and the sign of the intervening line of sight magnetic field. Notice the convention that $\mathrm{RM} >0$ when $B_{\rm l.o.s.}$ points towards the observer.

A striking theme emerging over the past decade is the presence of a seemingly universal toroidal magnetic field configuration in a wide variety of astrophysical systems.   Evidence from parsec- and kiloparsec-scale jets indicates that transverse RM gradients consistently correspond to outward-flowing axial electric currents
\citep{Gabuzda2012,Christodoulou2016}.
Analogous structures have been inferred in radio galaxies such as M87, 3C303, and Centaurus A, as well as in the center, the outer disk and halo of the Milky Way \citep{Feain2009,Pshirkov2011,Law2011,Oppermann2015}.
In each case, the implied direction of the large-scale current is away from the center of the galaxy or the jet base.

Detecting this pattern in galactic halos has proven more difficult.
The low density of the halo plasma requires stacking or population-level
statistics \citep[e.g.][]{Wiegert2015}, and the local halo environment is often disturbed by star formation, feedback, or tidal interactions.
Nevertheless, large-scale outflows can transport centrally generated magnetic flux to halo radii. If the underlying field is produced by a mechanism with a characteristic polarity pattern such as the Cosmic Battery then its imprint may still survive with stacking.

Motivated by this possibility, we aim to determine whether disk galaxies host a net axial electric current flowing away from their centers as predicted by the Cosmic Battery. This is equivalent to establishing the prevailing sign of the toroidal magnetic field above and below the disk plane. Our earlier attempt to address this question using the CHANG-ES sample of 35 edge-on galaxies observed with the VLA did not yield a statistically robust detection \citep{MyserlisContopoulos}. The limitation was the innermost agitated galactic environment which perturbs any underlying large scale magnetic field structure. In the present paper we decided to investigate the magnetic field structure at larger scales (on the order of few hundred kpc) where the galactic environment may not be agitated as much, but the magnetic field and the electron density are much weaker (halo RMs on the order of a few rad/m$^{2}$ compared to galactic RMs on the order of several hundred rad/m$^{2}$). There, we are limited by the number density of background radio polarized sources in the sky that may be used to probe the sign of magnetic field gradients in galactic halos.

The advent of the Polarization Sky Survey of the Universe’s Magnetism (POSSUM) overcomes this obstacle.  POSSUM is a large-area broadband polarimetric survey conducted with the Australian SKA Pathfinder (ASKAP) telescope. POSSUM is designed to map the magnetized Universe across the southern sky by producing a dense grid of RMs of approximately 30--50 compact extragalactic radio sources per square degree \citep{Gaensler2025,ASKAPpossum2025}. With its far higher RM-source density, POSSUM enables a much more sensitive search for systematic patterns in the halo RM field of the average edge-on galaxy obtained through population stacking.

In section~2 below, we describe the universal magnetic field gradient pattern predicted by the Cosmic Battery. Then, in section~3, we show that indeed, the stacked POSSUM RM observations exhibit a consistent large-scale axial current pointing outward from the galactic center in both hemispheres. We confirm that this result does not arise from sample-selection biases. We conclude in section~4 with a short discussion on the broader implications of our result for the origin and evolution of magnetic fields in disk galaxies.

\begin{figure}[t]
    \centering
    \includegraphics[width=0.95\linewidth]{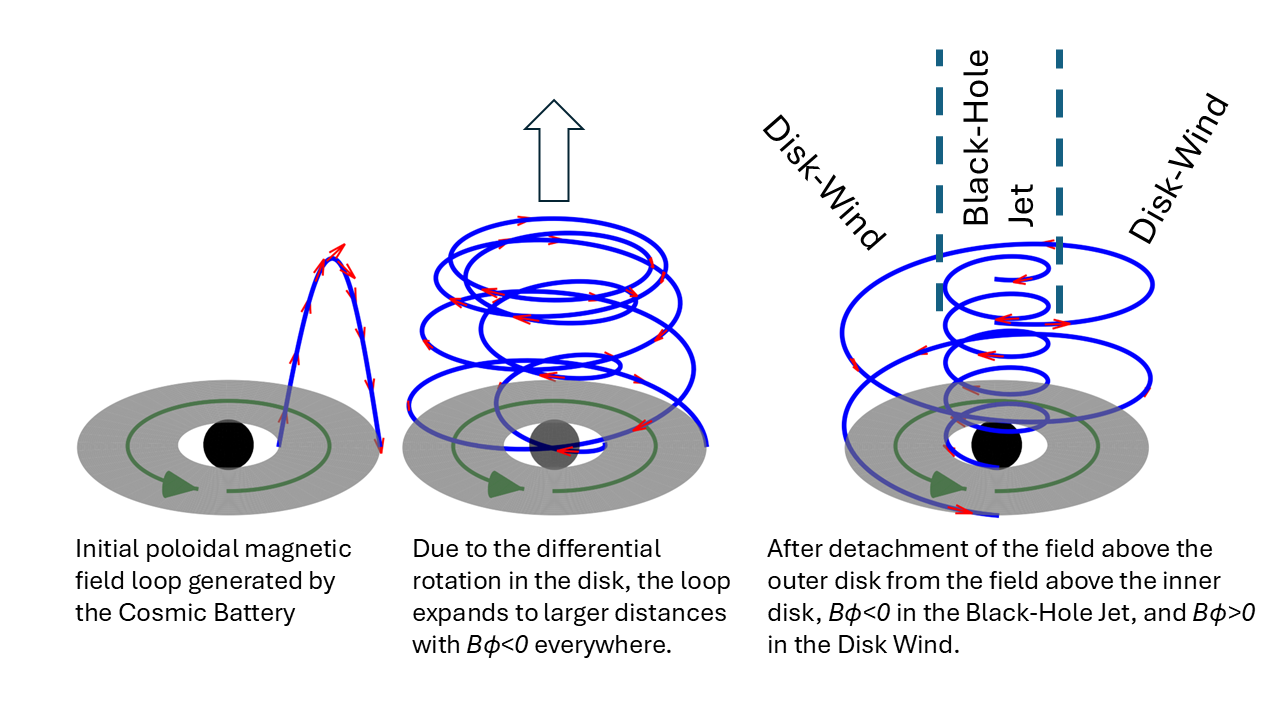}
    \caption{Cartoon showing the different directions of the toroidal $B_\phi$ field in the  axial Magnetic Black-Hole Jet and the outer Magnetic Disk-Wind predicted by the Cosmic Battery above (and below) the disk. Both configurations are based on an initial poloidal magnetic field loop generated by the Cosmic Battery configuration with $B_z\parallel \Omega$ in the inner part of the magnetic loop, and $B_z\parallel -\Omega$ in the return part of the magnetic loop in the outer disk. The azimuthal winding of the initial loop opens it up into the configuration of inner and outer magnetic winds. The green arrow shows the direction of disk rotation. }
    \label{1}
\end{figure}

\section{The prediction of a Universal Pattern}
\label{Cosmic_Battery}

The Cosmic Battery proposes that the magnetic field that threads the central supermassive black hole horizon is generated by the Poynting-Robertson drag effect on the plasma electrons \citep{ContopoulosKazanas1998,Contopoulos2018}. This yields a magnetic field structure where the axial magnetic field $B_z$ that threads the central black hole is aligned with the disk rotation, namely $B_z\parallel \Omega$ (we have aligned here the $z$-axis with the direction of $\Omega$). Several  more interesting conclusions are obtained from this result (see figure~1):
\begin{enumerate}
\item The magnetic field is generated in the form of magnetic loops that connect the inner and outer disk. 
\item $B_z$ in the innermost disk is aligned with the disk rotation, while $B_z$ in the outer disk is anti-aligned. This is very different from MHD simulations in which the polarity of $B_z$ is unrelated to the disk's direction of rotation.
\item The magnetic field loop above the disk are wound in the azimuthal ($\phi$) direction by the disk differential rotation, and thus a negative azimuthal component $B_\phi <0$ develops initially above the disk (similarly, $B_\phi >0$ below the disk). As the azimuthal winding continues, the whole loop expands and rises above and below the disk.
\item This configuration will not last for ever because once the magnetic loop expands sufficiently, the footpoint of the loop in the outer disk `forgets' that it was connected to the inner part of the loop, and the magnetic field above it will be bent {\it backwards} with respect to its direction of rotation. This is standard in MHD winds and jets from accretion disks, rotating black holes and pulsars. As a consequence, $B_\phi>0$ above the disk (and $B_\phi <0$ below), while $B_\phi<0$ above the black-hole (and $B_\phi >0$ below). 
\item The Black-Hole Jet configuration with $B_\phi<0$ is pushed to very large distances where it will form a shock. Beyond the shock, it will generate a backflow that will envelop the Magnetized Disk-Wind in which $B_\phi>0$.
\end{enumerate}

The innermost part of the magnetic field loop along the rotation axis will contain an electron-positron plasma generated by pair discharges in the charge-starved black hole
magnetosphere \citep[e.g.][]{Levinson2018}. This means that wherever the Black-Hole Jet is formed it will not generate any contribution to the RM of background sources crossing that region. Beyond some distance, however, the jet will mix with the outside plasma, it will form a shock and a backflow where it will contribute to the RM of background sources with polarity opposite to that of the Magnetized Disk-Wind. 

The reader may convince him/herself that a configuration with $B_\phi>0$ above the disk (and $B_\phi<0$ below the disk) always leads to a counter-clockwise  gradient of the RM across the axis of the galaxy. We may understand this by placing the galaxy horizontally with its $z$-axis vertically upwards. Then North is up, South is down, East is left, and West is right. $B_\phi>0$ in the North hemisphere implies that the RM contribution of the galactic $B_\phi$ is positive in the NE and negative in the NW quadrants. At the same time, $B_\phi<0$ in the South hemisphere implies that the RM contribution of the galactic $B_\phi$ is positive in the SW and negative in the SE quadrants. Therefore, all RM gradients across the axis of the galaxy are counter-clockwise. On the other hand, if we place the galactic $z$-axis vertically downwards with North again up, South down, East left, and West right, the RM contribution of the galactic $B_\phi$  is again positive in the NE and SW quadrants, and negative in the NW and SE quadrants. Therefore, all RM gradients across the axis of the galaxy are counter-clockwise irrespective of the galaxy orientation on the sky, and irrespective of its direction of rotation. An equivalent way to see this universality is that the electric currents inferred to flow along the galactic axis always flow from the center outwards in both hemispheres.

This universality of the gradients yields a particular pattern in the sky that resembles the sign of a well known German automotive firm. Such a universal pattern develops because of the connection between the direction of the axial field $B_z$ and the direction of rotation $\Omega$. If we change the direction of rotation, $B_z$ will change direction, but then this opposite magnetic field will be wound in the opposite direction than before, hence the direction of $B_\phi$ will be the same in all cases. This universality is not present in the standard dynamo scenario for the origin of the magnetic field, therefore, if confirmed, it will yield strong support to our Cosmic Battery scenario.

\section{The Universal Pattern observed with POSSUM}
\label{Analysis}

In order to investigate the universal pattern of toroidal magnetic fields generated by the Cosmic Battery around astrophysical black holes we decided to perform stacking of all available POSSUM background RM sources around large edge-on intervening galaxies. For that reason, we investigated a region of about 15 times the size of the galaxy which for galaxies in the POSSUM Southern hemisphere observing field turned out to be less than about $2^\circ$ around the larger galaxies in the field (of apparent size around $10'$). We used the following sample selection criteria:
\begin{enumerate}
\item Galactic disk inclination angle $>75^\circ$ (same as in the CHANG-ES edge-on galaxy sample). 
\item Galaxy apparent size in the sky between $6'$ and $10'$.
\item Region of interest with POSSUM background RM sources $2^\circ$ around each galaxy.
\item At least 2 background RM sources in each galactic quadrant for all galaxies.
\end{enumerate}

With these limits, we were able to collect the 10 galaxies in the POSSUM field listed in Table~A.1. We then implemented the following procedure:
\begin{enumerate}
    \item We resized all galaxies to a common size of 10', and rescaled accordingly the distances of all background RM sources found in the vicinity of that galaxy.
    \item We divided the sky around each resized galaxy into 4 quadrants around the galactic axis, and each quadrant into a near and far region beyond some separator distance $r_{\rm sep}$.
    \item We took the average RM from background sources in each near and far galactic quadrant for all galaxies. 
    \item We demeaned each quadrant average RM value based on the mean of all quadrant average RM values in the near region, and independently in the far region.
    \item We rotated all galaxies and their associated background RM sources horizontally. This rotation may be implemented within an extra rotation of $180^\circ$. Instead of randomly rotating multiple times each galaxy in the sample, we considered both horizontal configurations for each galaxy in the sample.
    \item Finally, we took the averages of the near and far quadrant RM averages for all galaxies and their horizontally flipped counterparts.   
\end{enumerate}

We notice that the obtained average values are much smaller than the standard deviation of background RM values in each wedge (a few rad~m$^{-2}$ vs few hundred), but they follow a very consistent pattern as shown in figure~2. This pattern may be considered as due to the toroidal magnetic field around the axis of the galaxy. The convention here is that positive points towards us, negative points away from us. Galactic Magnetic Winds around the axis of the galaxy carry one orientation of the toroidal magnetic field, while the axial Black-Hole Jet that operated during some stage of the galaxy's history carries the opposite orientation to even larger distance. If we associate the apparent galaxy major axis size of $10'$ with a typical galaxy size $2\times 10$~kpc, the Disk-Wind pattern manifests itself around the galaxy minor axis out to about 6 galactic radii or about $60$~kpc, while the Black-Hole Jet pattern manifests itself roughly beyond that distance. This is a new result that merits further investigation.

Our procedure described above was designed specifically to reveal any centrally symmetric underlying RM pattern that may be present in all galaxies. The averaging over each quadrant, near and far, was implemented in order for all galaxy quadrants to contribute equally in the final averaging. If a galaxy has less than 2 background RM sources in any one of its near and far quadrants, that galaxy is removed from the final averaging. The demeaning was applied so that the mean quadrant RM values in each near and far region of each galaxy fluctuate around zero. This procedure removes as much as possible RM contributions that are external to each galaxy. The $180^\circ$ horizontal reversal of each galaxy was implemented instead of implementing randomly multiple reversals in the sample. During the final averaging, this step `kills' non centrally symmetric RM patterns that may be present in any one galaxy and allows only
patterns with signs $(-,+,-,+)$ or $(+,-,+,-)$ in the (NW,NE,SE,SW) quadrants respectively. The final RM averaging in the near and far regions over all galaxies was applied to reveal which of the above two centrally symmetric underlying RM pattern, if any, is present in all galaxies. If no such universal underlying pattern existed, the final RM averaging would have yielded mean RM in all quadrants that would be consistent with zero (mean values smaller than one error of the mean). Instead, we observe the first pattern as expected in Magnetic Disk-Winds with the axial magnetic field configuration predicted by the Cosmic Battery inside about 6 galactic radii. Beyond 6 galactic radii we observe the second (reversed) pattern as expected in the Black-Hole Jet configuration due to the Cosmic Battery. The above are  shown in greater detail in figure~3 and in Table~A.2. The near pattern rapidly disappears beyond 6 galactic radii, and dissolves into the outer pattern beyond  roughly 10 galactic radii.

A per‑galaxy quadrant analysis shows that the inner sign pattern $(-,+,-,+)$ around the center is clearly present in 85\% of the sample galaxies, while the outer sign pattern $(+,-,+,-)$ is clearly present in 80\% of the sample galaxies. The full pattern in both the inner and outer regions is present in 50\% of all galaxies.
As we acknowledged above, the global signal is the sum of partial and very noisy contributions on the order of $\pm$ one hundred rad~m$^{-2}$. 
A leave‑one‑galaxy‑out analysis shows that removal of galaxies NGC2613 or ESO209-009 causes some degradation of the pattern, indicating that these systems contribute more significantly to the global signal. The above strongly argue against the pattern being a fluke from a single object. The observed quadrant asymmetry is a real but non‑uniform population effect, likely reflecting intrinsic differences in magnetic field coherence among galaxies. Notice that we did not implement any correction for the magnetic field of our own galaxy, nor for the magnetic field in the galaxy hosting each background RM source. We expect that the stacking of all galaxies in the extended POSSUM field and the implemented quadrant RM demeaning remove their RM contributions.

\begin{figure}[t]
    \centering
    \includegraphics[width=0.95\linewidth]{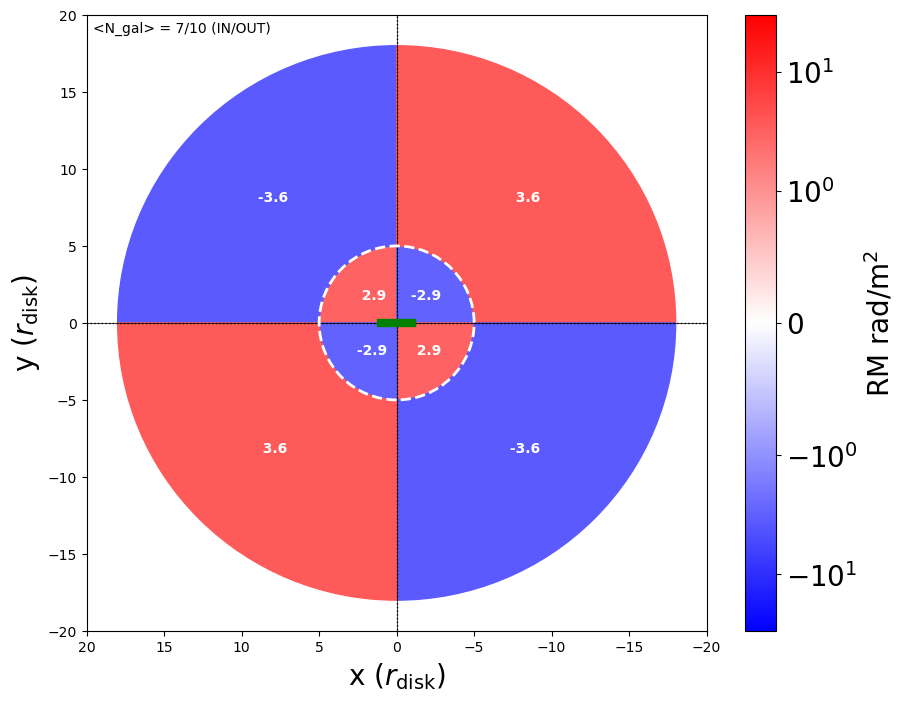}
    \caption{The observed universal pattern in the line-of-sight magnetic field around the average stacked edge-on galaxy for $r_{\rm sep}=5r_{\rm disk}$. In each near and far quadrant the average demeaned RM values are shown. This pattern is consistent with a magnetic field generated by the Cosmic Battery around the center of the galaxy. Inside $r_{\rm sep}$, the average pattern is $(-,+,-,+)$ in the (NW,NE,SE,SW) quadrants respectively and corresponds to a Disk Wind. Outside, the pattern is reversed to that of the Black-Hole Jet backflowing outside the Disk Wind.
}
    \label{2}
\end{figure}

\begin{figure}[t]
    \centering
    \includegraphics[width=1.00\linewidth]{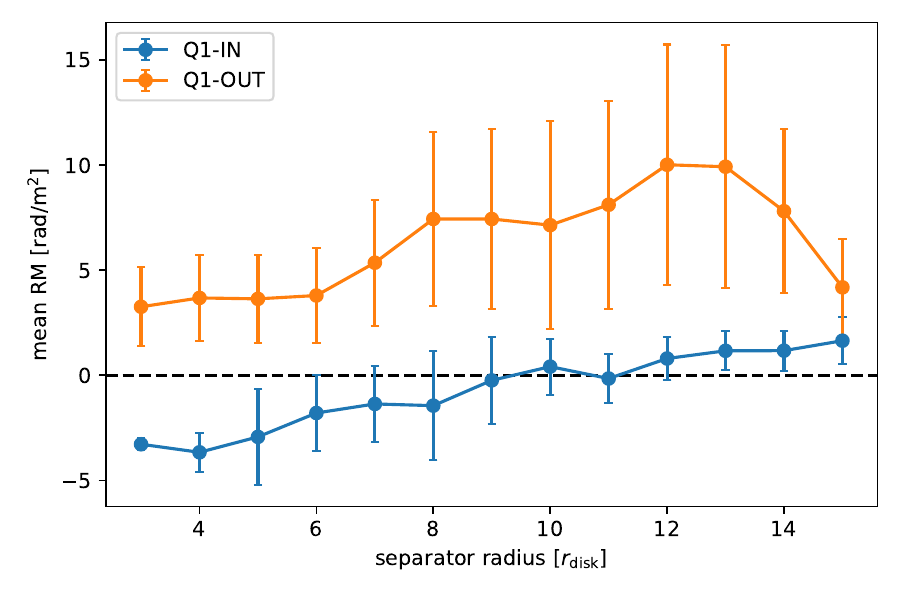}
    \caption{The observed universal pattern in the NW quadrant for the inner and outer regions, with the inner region ending at various distances separator radii $r_{\rm sep}$ as multiples of the central stacked galactic disk radius $r_{\rm disk}$ (blue and orange lines respectively). Shown also are the errors of the mean. We see that the inner pattern is present for separator radii smaller than about $6r_{\rm disk}$ where the average quadrant RM demeaned value is larger than its corresponding error of the mean. The outer reverse pattern is always present, at least out to distances of $15r_{\rm disk}$.
    }
    \label{3}
\end{figure}

\section{Conclusion}

The present work consists one of the very few known detections of the signature of the halo and the circumgalactic medium (CGM). Our preliminary evidence for a universal centrally-symmetric pattern in the galactic halo magnetic field is consistent with the evidence presented by Contopoulos, Kazanas and collaborators \citep{Gabuzda2012,Christodoulou2016}
for a universal toroidal magnetic field structure in astrophysical jets, namely that electric currents along kpc-scale astrophysical jets, whenever observed, always flow outwards. The same universal structure seems to be also present in the halo hemispheres above and below the galactic disk, and we believe it is observable through measurements of the gradients of background RM sources across the axis of galaxies observed edge-on. These measurements are very hard to obtain because the background RM values are on the order of a few hundred $\mbox{rad}\ \mbox{m}^{-2}$, while the signal we are trying to measure is on the order of $\pm$ a few $\mbox{rad}\ \mbox{m}^{-2}$. This is the reason the signal remained unobserved up to now. In our present work, we enhanced the signal via stacking of multiple galaxies and their associated background RM sources.

This universal centrally-symmetric pattern consists a Universal Symmetry-Breaking--USB in the distribution of large scale cosmic magnetic magnetic fields which cannot be accounted for by pure MHD processes\footnote{The equations of MHD are quadratic in the magnetic field $B$, and therefore, a reversal of $B$ everywhere does not affect the dynamics.}. A possible explanation for this USB is offered by the Cosmic Battery mechanism for the origin of astrophysical magnetic fields in the disk surrounding an astrophysical black hole that was proposed almost thirty years ago by \cite{ContopoulosKazanas1998}. 
The MRI–Parker mechanism \citep[e.g.][]{Machida2013} also produces a centrally-symmetric RM pattern. However, unlike the Cosmic Battery mechanism, the predicted sign distribution is intrinsically time-dependent depending on the orientation of the magnetic field emerging from the disk, thus, no systematic RM pattern is expected after ensemble averaging over a large enough galaxy sample.
A new element of our present work is that at distances beyond about 10 galactic radii, the magnetic field manifests the reverse Black-Hole Jet configuration (figure~1 offers an explanation). 
We hope that our preliminary results will stimulate a discussion on the origin of astrophysical magnetic fields in galactic halos.


{\tiny
\begin{spacing}{0.6}
\section*{\tiny Acknowledgements}
We acknowledge the usage of the HyperLeda database (\url{http://leda.univ-lyon1.fr}).
This scientific work uses data obtained from Inyarrimanha Ilgari Bundara / the Murchison Radio-astronomy Observatory. We acknowledge the Wajarri Yamaji People as the Traditional Owners and native title holders of the Observatory site. The Australian SKA Pathfinder is part of the Australia Telescope National Facility (https://ror.org/05qajvd42) which is managed by CSIRO. Operation of ASKAP is funded by the Australian Government with support from the National Collaborative Research Infrastructure Strategy. ASKAP uses the resources of the Pawsey Supercomputing Center. Establishment of ASKAP, the Murchison Radio-astronomy Observatory and the Pawsey Supercomputing Center are initiatives of the Australian Government, with support from the Government of Western Australia and the Science and Industry Endowment Fund. The POSSUM project (https://possum-survey.org) has been made possible through funding from the Australian Research Council, the Natural Sciences and Engineering Research Council of Canada, the Canada Research Chairs Program, and the Canada Foundation for Innovation. SM acknowledges support from grant PID2023-146372OB-I00 funded by MICIU/AEI/10.13039/501100011033 and by ERDF, EU.
\end{spacing}
}

\bibliographystyle{aa}
\bibliography{Literature.bib} 

%
%

\begin{appendix} 
\section{Statistical data}

We present here the results of our analysis. Table~A.1 lists the 10 galaxies that we selected from the POSSUM catalog based on the selection criteria of Section~3. Table~A.2 lists the average demeaned RM values of the POSSUM background sources per quadrant and their corresponding error of the mean in the near and far regions, for various values of the separator distance. In this calculation, all selected galaxies and their background RM sources are rotated horizontally a second time with an extra $180^\circ$ rotation.

\begin{table}
\caption{List of POSSUM edge-on galaxies in the sample}
\begin{tabular}{lccr}
Galaxy name & RA           & DEC         & $N_{\mathrm{RM}}$ \\
            & HH:MM:DD.DDD & DD:MM:SS.SS &  \\
\hline
\hline
ESO\,154$-$023 & 02:56:50.380 & $-$54:34:17.10 & 156 \\
NGC\,1448      & 03:44:31.915 & $-$44:38:41.38 & 151 \\
ESO\,209$-$009 & 07:58:15.009 & $-$49:51:05.08 & 172 \\
NGC\,2613      & 08:33:22.841 & $-$22:58:25.21 & 134 \\
NGC\,5084      & 13:20:16.923 & $-$21:49:39.28 &  88 \\
ESO\,270$-$017 & 13:34:47.300 & $-$45:32:51.00 & 212 \\
NGC\,5746      & 14:44:56.005 & $+$01:57:17.06 & 152 \\
ESO\,274$-$001 & 15:14:13.429 & $-$46:48:33.12 & 479 \\
IC\,5052       & 20:52:01.633 & $-$69:11:35.93 & 164 \\
NGC\,7090      & 21:36:28.865 & $-$54:33:26.35 & 164 \\ 
\hline 
Total & & & 1872
%

\end{tabular}
\end{table}

\begin{table}
\caption{Average demeaned RM values $({\rm RM}_{\rm NW})$ in the NW galactic quadrant in the near and far regions, and the corresponding errors of the mean for various separator radii $r_{\rm sep}$. According to the procedure described in section~3 above, ${\rm RM}_{\rm NW}={\rm RM}_{\rm SE}=-{\rm RM}_{\rm NE}=-{\rm RM}_{\rm SW}$. Shown also the number of galaxies $(\#)$ with at least 2 background RM values in all quadrants per sample.}
\begin{tabular}{c|llllll}
  &  near & & & far & & \\
$r_{\rm sep}$ &   ${\rm RM}_{\rm NW}$    & error &  $\#$ & ${\rm RM}_{\rm NW}$    & error & $\#$ \\
 $(r_{\rm disk})$ & $({\rm rad}\ {\rm m}^-2)$ & & & $({\rm rad}\ {\rm m}^-2)$ & &  \\
\hline\\
2 & -5.2 & 0 & 1 &+3.1 & 2 & 10\\
4 & -3.7 & 0.5 & 5 & +3.7 & 2 & 10\\
5 & -2.9 & 2.5 & 7 & +3.6 & 2 & 10\\
6 & -1.8 & 1.5 & 8 & +3.8 & 2 & 10\\
7 & -1.4 & 1.5 & 8 & +5.4 & 2.5 & 10\\
8 & -1.4 & 2.5 & 9 & +7.4 & 4 & 10\\
10 & +0.4 & 1 & 10 & +7.1 & 5 & 9 \\
12 & +0.8 & 0.5 & 10 & +10.0 & 6 & 7 \\
14 & +1.2 & 0.5 & 10 & +7.8 & 4 & 7 \\
16 & +2.8 & 0.5 & 10 & +2.2 & 1.5 & 1 
\end{tabular}
\end{table}

\end{appendix}

\end{document}